**Title: Modeling the dynamics of hypoxia inducible factor-1α (HIF-1α) within single cells and 3D cell culture systems**

**Authors:** Joseph Leedale[a], Anne Herrmann[b], James Bagnall[b,1], Andreas Fercher[c,2], Dmitri Papkovsky[c], Violaine Sée[b] and Rachel N Bearon[a*]

**Affiliations:**

[a] Department of Mathematical Sciences, University of Liverpool, L69 7ZL, UK

[b] Department of Biochemistry, Centre for Cell Imaging, Institute of Integrative Biology, University of Liverpool, L69 7ZB, UK

[c] Biochemistry and Cell Biology, University College, Cork,

[1] Present address: Faculty of Life Sciences, University of Manchester, M13 9PT, UK

[2] Present address: B. Braun Melsungen AG, Graz (Austria)

*To whom correspondence should be addressed:

Rachel Bearon, Dept. of Mathematical Sciences, Peach Street, University of Liverpool, Liverpool L69 7ZL, UK, rbearon@liverpool.ac.uk (T) +44 (0)151 794 4022

**Abstract:**

HIF (Hypoxia Inducible Factor) is an oxygen-regulated transcription factor that mediates the intracellular response to hypoxia in human cells. There is increasing evidence that cell signaling pathways encode temporal information, and thus cell fate may be determined by the dynamics of protein levels. We have developed a mathematical model to describe the transient dynamics of the HIF-1α protein measured in single cells subjected to hypoxic shock. The essential characteristics of these data are modeled with a system of differential equations describing the feedback inhibition between HIF-1α and Prolyl Hydroxylases (PHD) oxygen sensors. Heterogeneity in the single-cell data is accounted for through parameter variation in the model. We previously identified the PHD2 isoform as the main PHD responsible for controlling the HIF-1α transient response, and make here testable predictions regarding HIF-1α dynamics subject to repetitive hypoxic pulses. The model is further developed to describe the dynamics of HIF-1α in cells cultured as 3D spheroids, with oxygen dynamics parameterized using experimental measurements of oxygen within spheroids. We show that the dynamics of HIF-1α and transcriptional targets of HIF-1α display a non-monotone response to the oxygen dynamics. Specifically we demonstrate that the dynamic transient behavior of HIF-1α results in differential dynamics in transcriptional targets.

**Keywords:** Hypoxia, HIF (Hypoxia Inducible Factor), Negative feedback loop, Prolyl Hydroxylase, spheroids, mathematical modeling, oxygen nanoprobes.

## 1. Introduction

Oxygen homeostasis is crucial for the normal function and maintenance of respiring cells. The result of an insufficient supply of oxygen to the cell is hypoxia, a condition that plays a key role in a number of human pathologies. Hypoxia Inducible Factor (HIF) family members are transcription factors that mediate the intracellular hypoxic response. During hypoxia, HIF induces the transcription of a series of genes involved in diverse adaptive functions such as angiogenesis, glycolysis, cell proliferation and iron metabolism [1]. HIF is a heterodimer, comprised of an oxygen-regulated α subunit and a constitutive β subunit. Dimerization between the two subunits is necessary for DNA binding. In normoxic conditions prolyl hydroxylases (PHDs) catalyze the hydroxylation of HIFα, promoting its proteasomal destruction. Since PHDs require molecular oxygen in order to hydroxylate HIFα, in hypoxic conditions, hydroxylation is decreased. This results in HIFα stabilization and increased transcriptional activity. There is a feedback loop in this system as HIF also induces the transcription of the PHDs. This increase in PHD levels can compensate for the reduction of their activity when oxygen availability drops [2].

Mathematical models of biochemical networks have improved our understanding of biological phenomena. In particular, feedback models have provided insight into robust biological adaptation [3] and into dynamic oscillatory and pulsatile behavior. Examples of dynamic behavior include models of circadian rhythms [4], the cell cycle [5], and the dynamic behavior of key regulatory transcription factors [6, 7]. Cellular signaling pathways are typically very complex, and models must be designed to tackle the scientific questions being addressed subject to the available experimental data. Dynamic feedback models are typically expressed as systems of ordinary or stochastic differential equations, or hybrid combinations thereof. For example stochastic models may be necessary to investigate heterogeneity in single-cell imaging data [8], whereas minimal deterministic models may be sufficient to probe the dynamic properties of biological oscillators [9]. Calibrating or fitting the models to the data is a mathematically complex process, which again depends on the scientific questions being addressed and the experimental data available. Within the context of transcription factor pathways, a system perturbation by a given stimulus that produces an oscillatory or pulsatile response is particularly suitable for feedback modeling. Determining whether experimental data is sufficient to parameterize a model [10], and how to include parameter variability [11] are important concerns when developing dynamic feedback models.

Motivated by novel experimental data describing the dynamics of the HIF-1α protein at the level of single cells, we previously developed a simple mathematical model to capture the HIF-1α-PHD negative feedback [12]. This mathematical model was built based on live imaging experiments of single cells experiencing a single hypoxic transition. Parameters in the model were optimized through a combination of fitting to single-cell dynamic data, and through data collected from additional experiments. Whilst developing this model, it became apparent that it was necessary to explicitly distinguish the PHD isoforms and thus we developed an extended model which included the isoforms PHD1, 2 and 3. Using this extended model we previously identified PHD2 as the main PHD responsible for HIF-1α peak duration [12]. Here, in theory sections 3.1-3.3, and results section 4.1, we provide more extensive details of model development and calibration. In results section 4.2 we then provide additional validation of the model by comparing model predictions with new experimental data (described in methods section 2.1) investigating the response of cells to pulsatile hypoxic stimulation.

Whilst our previous work focused on single cells experiencing a single hypoxic transition, in tissues cells respond to continuous temporal changes in oxygen as well as spatial gradients. For example, the vascularization and oxygenation status of solid tumors vary over time due to the dynamic process whereby new blood vessels are formed and sub-functional vessels collapse [13]. The transient cycles of hypoxia-reoxygenation (intermittent hypoxia) that are known to occur in solid tumors is a poorly appreciated therapeutic problem and it is associated with resistance to radiation therapy and impaired delivery of chemotherapeutic agents [14]. Moreover, cells will experience different levels of oxygenation, depending on their proximity to blood vessels. These can also vary over time due to cell migration within a tissue. Hence, due to a combination of spatial and temporal factors, cells will experience constant and complex changes in their oxygenation.

The results presented in this current paper explore the implications of HIF-1α temporal dynamics within a spatial setting. In order to achieve this, we used experimental data which measured the oxygen concentration within a 3D system of cancerous neuroblastoma cells forming tumorspheres (described in methods section 2.2). These tumorspheres were subjected to different degrees of hypoxia and oxygen concentration distributions within the spheres were measured using phosphorescent Pt-porphyrin based oxygen nanosensors [15, 16]. These data were used to parameterize a spatial reaction-diffusion model for oxygen concentration (theory section 3.4 and results section 4.3), which was then coupled to the intracellular model of the HIF-1α-PHD feedback loop (theory section 3.5). By explicitly incorporating the dynamic behavior of HIF-1α, instead of assuming that the HIF-1α concentration takes an equilibrium value dependent on oxygen concentration, we uncovered unexpected HIF-1α and PHD spatio-temporal dynamics. Specifically our model predicts that in a tumorsphere, HIF-1α levels may display overshoot dynamics near the surface even though the absolute oxygen levels may be higher at the surface than in the center (results section 4.4). We also used the model to predict how the proteins PHD2 and PHD3, which are transcriptional targets of HIF-1, display different temporal dynamics at different spatial locations in the tumorsphere (results section 4.5).

## 2. Material and methods
### 2.1. Single-cell imaging of HIF-1α and ODD

Cell culture and hypoxic incubation:

HeLa cells were grown in Dulbecco's Modified Eagle's Medium (DMEM) supplemented with 10 % fetal calf serum (FCS) (v/v) and 1 % Non essential amino acids (v/v), at 37 °C, 5 % $CO_2$. Cells (between passages 8 to 20) were plated at $1\times10^5$ cells/ml. For imaging experiments, cells were plated in 35 mm glass bottom dishes (Greiner bio-one, UK). Hypoxic incubation was performed directly on the microscope stage equipped with a PeCon incubator with an $O_2$ controller. The ODD-EGFP HeLa cell line was generated by transduction of a HIV-ODD-EGFP-ires dTomato lentivirus as described in [12].

Time-lapse confocal microscopy:

Cells were incubated on the microscope stage at 37 °C, 5 % $CO_2$, 1 % or 20 % $O_2$ and observed by confocal microscopy using a Zeiss LSM510 with a Plan-apochromat 63X 1.3 NA oil immersion objective. Excitation of EGFP was performed using an argon ion laser at 488 nm. Emitted light was detected through a 505-550 nm bandpass filter from a 545 nm dichroic mirror. For time-lapse experiments mean fluorescence intensity was extracted and the fluorescence intensity was determined for each cell using CellTracker version 0.6 software [17]. Further details of the experimental protocol are provided in [12].

### 2.2. 3D cell culture and oxygen levels measurements

Tumorsphere formation and oxygen probe loading:

The cell line SK-N-AS (human S-type neuroblastoma) was grown in Minimal Essential Medium with Earle's salts plus 10 % FCS (v/v) and 1 % Non essential amino acids (v/v) and maintained in a humidified incubator at 37 °C, 5 % $CO_2$. For $O_2$ imaging MitoImage$^{TM}$ NanO2 probe (Luxcel Biosciences, Ireland) was added to the culture medium at a final concentration of 10 μg/ml and incubated for 6 h at 37 °C. After the incubation period, loaded cells were washed twice with PBS, trypsinized and resuspended in Neurobasal media supplemented with EGF (20 ng/ml), FGF2 (20 ng/ml), 2 % B27 (v/v), 1 % L-Glutamine (v/v), 1 % N2 supplement (v/v). A total of $1\times10^4$ cells/well was seeded in a 96-well U-Bottom plate (non-tissue culture treated) to allow sphere formation. After 24 h spheres were carefully transferred to a glass bottom dish (Ibidi, Germany) and Neurobasal media was added to a total volume of 2 ml.

FLIM measurements:

Oxygen levels in spheroids were imaged in FLIM mode (fluorescence lifetime imaging) on an inverted wide-field fluorescence microscope Axiovert 200 (Zeiss) equipped with 40x objective, and $CO_2/O_2$ climate control chamber (PeCon), pulsed excitation with a 390 nm LED and detection with gated CCD camera (both LaVision BioTec) and filter cube (390/40 nm excitation and 655/40 nm emission). More detail can be found in [18].

## 3. Theory/calculation
### 3.1. Two-component negative feedback model

We consider the following minimal feedback-model: HIF-1α ($x$) induces the transcription of PHD ($y$) at rate $k$ and HIF-1α ($x$) is degraded via PHD ($y$) dependent hydroxylation at a maximal rate $h$. To prevent elimination of HIF-1α ($x$) and unbounded growth of PHD ($y$) we further suppose HIF-1α ($x$) is produced through basal synthesis at rate $S$ and PHD ($y$) is degraded at rate $d$. The sensitivity to oxygen in this model is represented by the function $h(C)$, where $C$ is oxygen concentration, which is the rate at which PHD induces the degradation of HIF-1α. In preliminary work we described the hydroxylation by simple mass action, that is the rate at which HIF-1α ($x$) was removed through PHD ($y$) dependent hydroxylation was given by $hxy$. However when we attempted to fit the model to single-cell data obtained by time-lapse confocal microscopy we were unable to get good fits of the single-cell bell shaped data; in particular we were unable to obtain a large amplitude transient peak in HIF-1α concentration without a significant change in equilibrium levels between normoxic and hypoxic conditions. The bell-shaped data was of particular interest from a biological regulation point of view, as it is suggests a negative feedback loop, and so a requirement of our model was that it should capture this feature. We therefore included a saturating response to hydroxylation, thus introduce an additional parameter, $\gamma$, representing the HIF-1α hydroxylation threshold. Thus the minimal 2-component model is given as:

$$\frac{dx}{dt} = S - h(C)y\frac{x}{x+\gamma}$$
$$\frac{dy}{dt} = kx - dy$$
(1)

We also investigated models which incorporated a PHD precursor, representing for example PHD mRNA. However we obtained high degradation rates of mRNA, suggesting that the mRNA dynamics were operating on a fast timescale, thus justifying our assumption that the mRNA are at a quasi-steady state [19]. We note that standard linear stability-analysis [20] shows that the unique equilibrium solution to this system is stable.

### 3.2. Parameter optimization

The two-component model was fit to time-series data which measured the amount of HIF-1α ($x$) in individual cells. The dataset comprised two de-oxygenation experiments consisting of 17 and 22 cells respectively. To fit the model, an error function ($\chi$) was defined as the sum of squared residuals to indicate the difference between the solution to the ordinary differential equations (ODEs) and the experimental data. An optimization algorithm was written in Matlab R2011a to search for model parameters which minimized the error function. The built-in functions *ode45* and *fminsearch* were used for solving the ODEs and minimization respectively.

There were six parameters to optimize: four parameters which were fixed for each cell for the duration of the experiment ($S$, $k$, $\gamma$ and $d$) and two values for the hydroxylation rate ($h_N$ in normoxia, and $h_H$ in hypoxia) for each cell. As PHD ($y$) data was not available, if we attempted to fit the model to experimental data, then several parameters in the model would not be identifiable. For example the solution for $x(t)$ is unchanged if we replace $y(t)$ by any scalar multiple, say $cy(t)$, and also replace $k$ with $ck$ and $h$ with $h/c$. Therefore, the parameters $k$ and $h$ are unidentifiable. In order to obtain a model which we could fit to data, that is a model which was structurally identifiable, we chose to rescale $y$ with $h_N$, the hydroxylation rate of HIF-1α in normoxia. With this choice of rescaling, the parameter $k$ now represents the original $k$ multiplied by $h_N$, and $h = 1$ in normoxia and $h = h_H/h_N$ in hypoxia. Furthermore we took $h_H/h_N = 0.14$, based on measured values from the literature [21] for the hydroxylation rate of the PHD2 isoform (widely considered to be the main oxygen sensor, see e.g. [22]). The number of free parameters to optimize in the system was thus reduced to four.

For each cell dataset we generated 50 parameter sets $[k, \gamma, d, S]$ to use as initial values to find a global optimal parameter set which best fit each experimental data set. The parameters $d$ and $S$ were specified based on the experimental data, and the parameters $k, \gamma$ were sampled from a distribution. The half-life of the pathway's main oxygen sensor, PHD2, has been measured at 785 min [12]. This gave an initial estimate for $d$ of $8.83 \times 10^{-4}$ min$^{-1}$. The Michaelis constant $\gamma$ should be comparable to the amount of protein for which saturation is observed to justify its inclusion in the model, thus the sampling range for $\gamma$ was taken to be centered on $x(0)$, varying one order of magnitude either way. In preliminary fitting of a small subset of data, the value estimated for $k$ was very consistent, falling between $2.24 \times 10^{-4}$ and $5.49 \times 10^{-4}$ for 6 cells. The sampling range was centered on the average of these values varying one order of magnitude either way. We assumed that prior to the oxygen switch, $t < 0$, the system was at equilibrium $(x^{EQ}, y^{EQ})$ which satisfies

$$\begin{aligned}\frac{dx}{dt} &= S - hy^{EQ}\left(\frac{x^{EQ}}{x^{EQ} + \gamma}\right) = 0, \\ \frac{dy}{dt} &= kx^{EQ} - dy^{EQ} = 0.\end{aligned} \quad (2)$$

We took the first experimental data-point as an initial estimate for the value of HIF-1α at $t = 0$, i.e. $x(0)$. Initial estimates for the values of $y(0)$ and $S$ were then computed by assuming equilibrium at $t = 0$:

$$y(0) = \frac{kx(0)}{d} \quad (3)$$

$$S = h\frac{kx(0)}{d}\left(\frac{x(0)}{x(0) + \gamma}\right) \quad (4)$$

### 3.2.1. Free optimization of bell-shaped data

The model parameters were initially optimized by separately fitting the model to each of 11 cells that shared a qualitative bell-shaped property in their transient HIF-1α dynamics during hypoxic induction. This dataset was chosen as it motivated the development of the minimal model described above. Furthermore, we anticipated that the data from each cell would correspond to a unique optimal parameter set; that is for each data set the model parameters would be identifiable. In contrast, we anticipated that the model could adequately fit each data set showing simpler dynamics for several choices of parameters; that is the model parameters would be unidentifiable.

### 3.2.2. Constrained optimization

All heterogeneity in the experimental data was accounted for in our optimization process through the variation of parameters. By examining the experimental protocol and the model we could predict which parameter values may be expected to vary between cells. Taking all the experimental data alongside the model, these parameters could be interpreted as being globally unidentifiable; that is whilst we anticipated that they would be identifiable for each bell-shaped experimental data, we did not expect to be able to identify a unique value across all the experimental data. Parameters which were not expected to vary between cells, or at least only vary due to endogenous cell-to-cell heterogeneity, were interpreted as being globally identifiable. We undertook a parameter search where globally identifiable parameters were constrained with the aim of obtaining a unique optimal parameter set for each and every cell data set.

The dynamics in the experimental data are represented by arbitrary units (A.U.) of fluorescence produced as a result of the transient transfection of the cells with HIF-1α green fluorescent fusion protein. These fluorescence units are not comparable between separate experiments or between cells in the same experiment for at least two reasons. The method of transient transfection leads to cells with different copy numbers of the HIF-1α fusion constructs. In our model this corresponds to $S$, the basal synthesis rate of HIF-1α, not being conserved across cells. Also, the experimental imaging protocol involves manually adjusting laser settings in order to optimize the image. Gain and offset values are fine-tuned at the start of each experiment to minimize noise-to-signal ratio and avoid saturation of the signal. Thus fluorescence units between independent experiments cannot necessarily be directly compared. For details on how these factors affect parameter variation see the Appendix.

We again initialized our search using 50 parameter sets $[k, \gamma, d, S]$. However, in the constrained optimization, we fixed the initial estimates of $k$ and $d$ to be the median values found in the free optimization of bell-shaped data (section 3.2.1). Initial values of $S$ and $\gamma$ were chosen as described above.

In the constrained parameter optimization, estimates for $k$ and $d$ were constrained to within 50 % of median values found in the free optimization of bell-shaped data. Converged solutions were further classified as good or bad fits using an error envelope. Specifically, we constructed an error envelope with upper and lower bound defined by:

$$\text{EXP}(t) \pm \beta\big(\max(\text{EXP}(t)) - \min(\text{EXP}(t))\big) \qquad (5)$$

where $\text{EXP}(t)$ represents the time-series vector of experimental data and $\beta$ defines the width of the envelope, which was taken as $\beta = 0.35$. Solutions were classified as bad fits if more than 1 % of the experimental data points lay outside the error envelope and good fits otherwise. The values for the envelope width, $\beta$, and the deviation from the envelope provided an objective measure of goodness of fit but were chosen arbitrarily.

### 3.3. Extension of model to four components

The minimal two-component ODE model was extended to four components by removing the generic PHD feedback variable ($y$) and explicitly accounting for the three different isoforms of PHD: PHD1, PHD2 and PHD3 (termed $y_1$, $y_2$ and $y_3$ respectively):

$$\begin{aligned}
\frac{dx}{dt} &= S - \frac{x}{x+\gamma} \sum_{i=1}^{3} h_i(C) y_i \\
\frac{dy_1}{dt} &= S_1 - d_1 y_1 \\
\frac{dy_2}{dt} &= S_2 + kx - d_2 y_2 \\
\frac{dy_3}{dt} &= S_3 + kx - d_3 y_3
\end{aligned} \qquad (6)$$

Biological justification for the model development is provided in [12]. In summary, we take PHD2 and PHD3 to be HIF-1α-inducible with equal induction rate $k$, whilst PHD1 is not inducible. For $i = 1$ to 3 basal synthesis and degradation rates for PHD$i$ are given by $d_i$, and $S_i$; and PHD$i$ causes hydroxylation of HIF-1α at a rate $h_i$ which is dependent on oxygen concentration, $C$. Specifically, the equations for the hydroxylation rates, normalized on $h_2$ in normoxia ($C = 20$ %), are based on data from [21] and given by:

$$\begin{aligned}
h_1(C) &= 0.25(0.0014 C^2 + 0.016 C + 0.1233) \\
h_2(C) &= 0.0015 C^2 + 0.0137 C + 0.1202 \\
h_3(C) &= 1.25(0.0022 C^2 + 0.0012 C + 0.1036)
\end{aligned} \qquad (7)$$

The induction rate, $k$, and synthesis rate of one of the PHDs were optimized to obtain a best fit to the 2-component model with median parameter values. Other parameters were constrained by experimental data as detailed in [12].

### 3.4. Model for oxygen concentration

We model the tumorsphere as a radially symmetric sphere of radius $R$ and assume oxygen diffuses at rate $D_1$ inside the sphere and rate $D_2$ outside the sphere. Within the sphere, we assume cells consume oxygen. Thus the oxygen concentration, $C(r)$, satisfies the following equation:

$$\frac{\partial C}{\partial t} = \begin{cases} \frac{D_1}{r^2}\frac{\partial}{\partial r}\left(r^2\frac{\partial C}{\partial r}\right) - \psi C, & r < R, \\ \frac{D_2}{r^2}\frac{\partial}{\partial r}\left(r^2\frac{\partial C}{\partial r}\right), & r > R, \end{cases} \quad (8)$$

subject to boundary conditions:

$$\frac{\partial C}{\partial r} = 0, \quad r = 0, \quad (9)$$

$$C \to C_\infty, \quad r \to \infty, \quad (10)$$

$$D_1 \frac{\partial C_-}{\partial r} = D_2 \frac{\partial C_+}{\partial r} \quad \text{at } r = R. \quad (11)$$

The first boundary condition ensures no singularities occur at the origin and the second boundary condition specifies the far-field oxygen-concentration, $C_\infty$. The final boundary condition matches the diffusive flux at the surface of the sphere to ensure conservation of mass at the surface.

### 3.4.1. Steady state solution and parameter fitting

Non-dimensionalizing radial position on $R$, by using standard techniques for solving ODEs (e.g. [23]) we find the steady-state solution is given by

$$C = \begin{cases} \dfrac{C_0 \sinh(\sqrt{\tilde{\psi}}r)}{r \sinh(\sqrt{\tilde{\psi}})}, & r < 1, \\ \dfrac{1}{r}(C_0 - C_\infty) + C_\infty, & r > 1, \end{cases} \quad (12)$$

where $\tilde{\psi}$ is the uptake rate non-dimensionalized on $R^2/D_1$ and the concentration at the surface of the sphere, $C_0$, is given by

$$C_0 = \frac{D C_\infty}{\sqrt{\tilde{\psi}} \coth\left(\sqrt{\tilde{\psi}}\right) - 1 + D} \quad \text{where } D = \frac{D_2}{D_1}. \quad (13)$$

To estimate the model parameters, we simultaneously fitted the steady-state solution to four sets of experimental data corresponding to 2 cross-sections of a tumorsphere for two specified values of $C_\infty$. Given the radius, $R$, was known, we optimized the quantity $\Psi = \tilde{\psi}/R^2 = \psi/D_1$ which should be the same for all data regardless of sphere radius. We also optimized the constant parameter $D$.

### 3.4.2. Unsteady numerical solution

Initially, the unsteady oxygen concentration was computed numerically with time non-dimensionalized on the timescale for diffusion across the sphere, $R^2/D_1$. Solutions to the time-dependent system were found numerically using a backward-time centered-space (BTCS) implicit Euler finite difference method.

To couple the unsteady solution for the oxygen dynamics with the HIF-1α model, we required an estimate of the diffusion rate inside the sphere, $D_1$. This was necessary in order to run the oxygen model on the same timescale as the HIF-1α model. Whilst detailed information regarding the temporal oxygen dynamics was lacking, experiments have shown that it takes 5 min for a tumorsphere of radius 187.5 μm to attain oxygen equilibrium when subject to an oxygen switch from 8 % to 3 % (data not shown). We could then estimate $D_1 = R^2 T / t_{eq}$, where $R = 187.5$ μm, $t_{eq} = 5$ min, and $T$ is the non-dimensional time taken in the numerical simulation for equilibrium to be attained after a sphere is subject to an oxygen switch from 8 % to 3 %. We defined the numerical simulation to be at equilibrium if the relative error between the oxygen concentration and the analytic steady-state was less than 1 % at every grid point.

### 3.5. Coupling spatial oxygen model with HIF-1α model

The oxygen spatial model was coupled to the HIF-1α model by computing the temporal dynamics of both the oxygen and HIF-1α at different spatial locations. In this simple model, we neglected details of cell movement and growth, and assumed oxygen uptake is uniform throughout the tumorsphere. We thus did not take explicit account of individual cells, and instead focussed on how the HIF-1α dynamics would vary spatially due to the spatiotemporal oxygen signal. At a particular spatial location, we could compute the oxygen temporal dynamics which were then input into the hydroxylation function (equations (7)) used in the HIF-1α model. Specifically, the model for the unsteady oxygen dynamics was solved to obtain $C(r_i, t_j)$ for $i = 1..N$, $j = 1..M$, where $r$ and $t$ are equally spaced vectors representing radial position and time with $r_1 = 0$, $r_N = 10$, $t_1 = 0$, $t_N = 30$ h. Note that space is non-dimensionalized on the sphere radius, whilst time is dimensional in order to couple with the dimensional HIF-1α model. The grid values taken were $N = M = 1000$, and convergence of the solution was checked by increasing the spatial and temporal resolution. The HIF-1α ODE model as described in section 3.3 was then solved for $r_i$, $i = 1..N$ by taking temporally varying hydroxylation rates (see equations (7)) based on the oxygen concentration given by $C(r_i, t_j)$. Conversion between the internal oxygen concentration given in units of μM to equivalent external oxygen concentration given as % volume as measured in the atmosphere of the hypoxic chamber was performed using previous calibration results [18, 24].

### 4. Results and Discussion
#### 4.1. A minimal 2-component model for HIF-1α-PHD negative feedback can capture dynamic single-cell data.

Numerical simulations of model (1) were compared with experimental data [12]. By varying parameters in the model, we previously obtained a good fit to the experimental bell-shaped data (Fig 1, adapted from [12]). Furthermore, in most cases, the optimal parameters were independent of initial estimates for the parameters, thus suggesting that we have found a unique global optimal set. Specifically, Fig 2 shows the results of the parameter optimization, in which 50 initial sets of parameter were randomly selected for each cell. For all cells, at least 32 of the initial sets converged to a $\chi$ value (measure of error between numerical solution and experimental data) within 1 % of the minimum $\chi$ value. The optimal parameters generally converged to a unique parameter set, although we note exceptions, for example the value of $S$ in cell 9 displays variability depending on the initial parameter estimates. Much greater variability was found when applying the free optimization method to non-bell-shaped data. For example, multiple parameter sets were equally good at fitting experimental data which shows a simple linear increase in HIF-1α (data not shown).

Under free optimization, we see large variability in optimal parameters across the cells (Fig 2). However we would expect the parameters $k$ and $d$ to be conserved (see section 3.2.2 and [12]). In Fig 3, the optimal parameters were obtained via a constrained optimization for the full set of experimental data. In the constrained search, parameters $k$ and $d$ were constrained to within 50 % of the median values depicted in Fig 2. For each cell, at least 42 of the initial sets converged to a $\chi$ value within 1 % of the minimum $\chi$ value. The model could successfully fit 31 out of 39 cells according to the error envelope criteria. Fig S1, Fig S2, Fig S3 provide additional plots of all the simulations fit to the experimental data, and identify cells which fail the error envelope criteria.

In Fig 3 we see that the bell data in *Experiment 2* had higher values of $S$ and lower values of $\gamma$ than data from *Experiment 1*. Whereas imaging protocols would predict a positive correlation between measured values of $S$ and $\gamma$, variation in HIF-1α copy number between cells as a result of transient transfection may lead to variation in $S$ but not $\gamma$ (see scaling analysis of section 3.2.2). Therefore it could be that $\gamma$ values are generally lower in *Experiment 2* than *Experiment 1* as a result of the imaging scaling but $S$ values are higher due to a significantly higher copy number of HIF-1α in *Experiment 2* as a result of transient transfection. This hypothesis may also be linked to why we only see bells in *Experiment 2*. Alternatively, the exclusivity of bells to *Experiment 2* may suggest that there are other means of regulation (e.g. oxygen independent) involved that we have not accounted for in the model. This could include for example environmental factors such as Fe2+, 2-oxoglutarate (2OG) and ascorbate.

### 4.2. Predictions and experimental validation of the extended model

As demonstrated in [12], the extended 4-component model, which is more tightly constrained by additional experimental measurements, fits to the 2-component model quite well. We also have previously demonstrated that removal of PHD2 leads to sustained high-levels of HIF-1α, indicating that PHD2 is the main PHD responsible for controlling HIF-1α dynamics. Here we investigate the effect of repetitive hypoxic shocks applied to the system through model prediction (Fig 4A) and experimental validation (Fig 4B). Both the prediction and validation show a reduction of the amplitude of HIF-1α response for the second peak. This occurs because PHD2 and PHD3 are not reduced to basal equilibrium levels during the single hour of normoxia which separates the hypoxic shocks.

### 4.3. The oxygen dynamics in a tumorsphere can be captured by a diffusion model

Steady state experimental measurements of oxygen concentration across the diameter of a tumorsphere in normoxia and hypoxia are shown in Fig 5. Also shown are steady-state solutions to a radially-symmetric reaction-diffusion equation representing oxygen uptake (respiration) within the tumorsphere and diffusion both inside and outside the tumorsphere. The model is fit to the experimental data by adjusting only two parameters: $\Psi = \psi/D_1$, the ratio of oxygen uptake to oxygen diffusion inside the sphere and $D = D_2/D_1$, the ratio of diffusion outside to inside the sphere. Estimates for these parameters are $\Psi = 3.35 \times 10^{-4}$ μm$^{-2}$ and $D = 3.48$.

We obtained an estimate for the non-dimensional time to attain equilibrium to be $T = 13.95$ which, based on the experimental estimate of 5 min for $R = 187.5^2$ μm, leads to an estimate of internal diffusion rate of $D_1 = R^2 T/t_{eq} = 1.63 \times 10^{-5}$ cm$^2$ s$^{-1}$. This appears a plausible estimate when compared to a typical diffusion rate of oxygen in water, 2.5×10$^{-5}$ cm$^2$ s$^{-1}$ [25]. We can also compare our estimate for oxygen consumption rate with literature values. We estimate the uptake rate to be $\psi = 3.29 \times 10^1$ min$^{-1}$ which for a cell of volume 5×10$^3$ μm$^3$ gives an uptake rate per cell ranging from 2.8×10$^{-17}$ mol s$^{-1}$ at the centre of the sphere (corresponding to oxygen concentration of 10 μM) to 11×10$^{-17}$ mol s$^{-1}$ at the outside (oxygen concentration of 40 μM) of the relatively normoxic sphere depicted in Fig 5. This is comparable to literature values for oxygen uptake in spheroids computed through measurements of external oxygen concentration, for example from $1 \times 10^{-17}$ mol cell$^{-1}$ s$^{-1}$ to $16 \times 10^{-17}$ mol cell$^{-1}$ s$^{-1}$ [26, 27].

### 4.4. Hypoxic shock causes HIF-1α overshoot dynamics near surface of tumorsphere

Results of a coupled numerical solution of the extended HIF-1α dynamic model with unsteady reaction-diffusion model for oxygen dynamics are shown in Fig 6. Initially, at normoxic equilibrium, HIF-1α levels are elevated towards the center of the sphere where oxygen levels are depleted due to oxygen uptake by the cells and diffusion-limited transport through the outer boundary of the sphere. When a tumorsphere experiences a rapid change to hypoxia, the oxygen-level at the boundary drops and HIF-1α levels rise rapidly. Because the transition in oxygen is most rapid at the boundary, the transient HIF-1α dynamics are also most significant at the outer boundary. Furthermore, although the oxygen concentration has attained a new equilibrium by $t = 2$ min, there is still significant dynamic behavior in the HIF-1α levels, which varies across the sphere. Once HIF-1α dynamics stabilize to a new equilibrium state (e.g. solution at $t = 30$ h) we again see HIF-1α levels inversely correlated with distance from the center of the sphere corresponding to the correlation between steady-state oxygen levels and distance from the center of sphere.

This result questions the conventional assumption that low oxygen levels equate to high levels of HIF-1α, and high oxygen levels equate to low levels of HIF-1α. Whilst the assumption is true when considering the equilibrium state; the dynamic response of HIF-1α to changes in oxygen levels can result in the assumption being false. Specifically we here show that the highest levels of HIF-1α are found at the surface of a tumorsphere where the oxygen levels are highest, because the surface is where the fastest change in oxygen occurs. We note that these results are based on an average HIF-1α temporal response, and we have neglected the observed cell-to-cell heterogeneity in the single-cell data. For example, if a large proportion of the cell population display a gradual increase in HIF-1α (e.g. see Fig S2) instead of the bell-shaped response, the surface overshoot dynamics might not be observed.

### 4.5. The dynamic transient behavior of HIF-1α results in differential dynamics in the expression of transcriptional targets.

In Fig 7, we model a tumorsphere that experiences an acute hypoxic switch from 20 % oxygen to hypoxia (1 %). As in Fig 6, we present results from the coupled numerical solution of the extended HIF-1α dynamic model with unsteady reaction-diffusion model for oxygen dynamics. However, in Fig 7 the boundary conditions represent a greater hypoxic shock than considered in the tumorsphere experiments depicted in Fig 5 and modeled in Fig 6. The response to oxygen is incorporated through the hydroxylation rates given by equations (7). In Fig 7, we present the dynamic response at different spatial locations of both HIF-1α and transcriptional targets PHD2 and PHD3 which appear in the extended HIF-1α dynamic model. We see a pronounced boundary overshoot for both HIF-1α and PHD3 levels following this severe hypoxic switch. In contrast, PHD2 levels are ordered such that there is always more PHD2 in more hypoxic regions. This is likely due to the significant difference of stability between PHD2 and PHD3 (measured in [12]). PHD3 has a shorter half-life and therefore responds more rapidly to the hypoxic switch. Our model thus predicts that the expression levels of the transcriptional targets of HIF-1, in this case the PHD isoforms, can display differential dynamics when subjected to the same hypoxic signal, due to HIF-1α dynamics.

This result highlights the implications of the dynamic HIF-1α model within a spatial context. We remind that these results are based on an average HIF-1α temporal response, and that cell-to-cell heterogeneity may lead to alternative patterns in the temporal-spatial dynamics of target genes. HIF-1α is a transcription factor for many target genes, and so correctly incorporating the dynamics of HIF-1α appropriate for a given spatial location is necessary to identify the dynamics of target genes. Identifying differential dynamics in target proteins may have implications regarding spatial variability in cell fate.

### 5. Conclusions

*In vivo*, cells do not experience hypoxia as a binary on/off switch; instead cells respond to a range of spatial and temporal gradients in oxygen, mediated through a dynamic signaling pathway. A key novelty was to demonstrate that the protein HIF-1α does not simply track oxygen levels, but displays transient overshoots in response to a hypoxic shock. By developing a model based on this single-cell experimental data, we have been able to predict how the dynamics of HIF-1α are affected by a range of spatial and temporal oxygen signals.

Alternative models investigating how HIF mediates the cell's response to hypoxia often have focused on how equilibrium levels of HIF are a function of oxygen levels [28-30]. In developing our model, we have attempted to ensure that the model is suited to answer the scientific questions we wanted to address, and have parameterized it using available data. In future, our model should be integrated with a recent model incorporating data on the transcriptional activity of HIF-1 [31]. In our model we have accounted for cell-to-cell heterogeneity in the data via parameter variation using optimization tools. We are aware that fitting dynamic data to models is an area of active research, and appreciate that the fitting of our model would likely benefit from an increased level of sophistication with stochastic and statistical tools [11]. We also note that whilst we were able to fit our two-component model to a range of dynamic responses, the later conclusions of the study made use of an average cell constructed using median parameter values. It would be interesting to understand more clearly the impact that cell-to-cell heterogeneity has on these later conclusions.

The applications of this model have more general use than just the study of the hypoxic response. Our mathematical model represents the evolution of a transcription factor that is predisposed to rapidly accumulate in an overshoot fashion, as the result of an acute change in signal. This qualitative overshoot characteristic may be lost in multi-cell environments or at least the rapidity of the peak transiency may be hidden among cells responding at slightly different times or gradients. Furthermore, these dynamics may be ignored completely in some signaling pathway models, especially if the transcription factor is an intermediate part in a long chain or the timescale of the phenomenon being studied is sufficiently separated from the transcription factor dynamics. These oversights could potentially result in the loss of any feedback information that propagates from these dynamics. This propagation could be in the form of downstream effects, which for transcription factors means target gene regulation. Our model is also useful in a general sense as it includes the description of target genes that have similar roles (negative feedback) but respond at different rates. This research is potentially applicable to other transcription factors that share transient dynamics as a result of acute signal changes and regulate the transcription of differently responding target genes.

In the context of tumorspheres, we have developed a simple model for oxygen diffusion and uptake, parameterized with oxygen data obtained from experimental intracellular probes. We predicted that cells on the surface of the sphere experience a more acute switch and thus are more likely to experience transient overshoot dynamics than cells within the center of the sphere. This effect can compensate for the initially higher oxygen levels, and consequent lower levels of HIF-1α, to result in transient dynamics that have higher HIF-1α levels at the surface of the sphere than in the central hypoxic region. This result suggests a potential mechanism for manipulating the signaling pathway to deliver a stronger hypoxic response (HIF-1α activity) in less hypoxic cellular regions. This manipulation could take the form of a specific oxygen signal while the response of interest might be governed by a downstream target gene. This hypothesis should be tested experimentally by simultaneously measuring oxygen and HIF-1α within tumorspheres. In future, a more sophisticated model may be necessary which includes effects such as uptake rate being dependent on oxygen concentration. We have been able to better fit the experimental data of Fig 5 by allowing uptake rate to be a saturating response (data not shown). However the increased mathematical complexity did not modify the conclusions presented here. We also note recent relevant work focusing on modeling the oxygen dynamics in spheroids with a distinct necrotic core [32]. In that work, non-constant oxygen consumption is considered, but the boundary conditions differ from that considered here. In future, developing and extending the model to more realistic tumor geometries and physiological oxygen dynamics would provide useful insights into *in vivo* HIF-1α dynamics.

HIF-1 directly induces the transcription of a multitude of genes, and also couples to many other signaling pathways, for example the p53 pathway [33]. We previously have shown the impact that dynamic HIF-1α has on a model for the p53 pathway, and investigated the impact of knocking out PHD2 [12]. Here, our model allowed us to predict how the dynamic behavior of HIF-1α in response to oxygen dynamics impacts the transcription of downstream genes within a spatial setting. As a specific example, we predicted the transient dynamics of PHD isoforms 2 and 3 within a sphere subject to a hypoxic shock. Because the timescales of degradation differed between the two isoforms, we were able to predict differential expression levels. For example, boundary cells experienced transient high levels of PHD3, but not PHD2. Coupling the spatial model to signaling pathways which cross-talk with HIF is crucial for making predictions as to how the body responds to hypoxia, for example in the response of tumors to medical intervention [34]. Moreover, considering the recent interest in targeting HIF for cancer treatment [35], a better understanding of its levels and activity with spatial and temporal resolution could potentially help to achieve more focused drug targeting.

**Appendix**

We fit our 2-component HIF-1α feedback model to experimental single-cell data generated using transient transfection and time-lapse confocal microscopy. To understand how experimental protocols affect parameter variation, we considered two sets of model variables and parameters: the *real* variables and parameters (superscript $R$); and the *measured* or *fitted* variables and parameters (superscript $F$). So the *real* system as represented by our model has the following form:

$$\frac{dx^R}{dt} = S^R - h^R y^R \left(\frac{x^R}{x^R + \gamma^R}\right)$$
$$\frac{dy^R}{dt} = k^R x^R - d^R y^R$$
(A1)

We assume that *measured* HIF-1α is related to *real* HIF-1α by a scaling factor $A$, and that *fitted* PHD is re-scaled on the maximal hydroxylation rate of the *measured* HIF-1α in normoxia:

$$x^R = A x^F \tag{A2}$$

$$y^R = \frac{A}{h_N^R} y^F \tag{A3}$$

We thus obtain:

$$\frac{dx^F}{dt} = \frac{S^R}{A} - \frac{h^R}{h_N^R} y^F \left( \frac{x^F}{x^F + \frac{\gamma^R}{A}} \right)$$

$$\frac{dy^F}{dt} = k^R h_N^R x^F - d^R y^F$$

(A4)

If we compare this to the *fitted* system:

$$\frac{dx^F}{dt} = S^F - h^F y^F \left( \frac{x^F}{x^F + \gamma^F} \right)$$

$$\frac{dy^F}{dt} = k^F x^F - d^F y^F$$

(A5)

we see that the *measured/fitted* HIF-1α basal synthesis and Michaelis constant, $S^F$ and $\gamma^F$, are equal to their real values divided by the scaling factor $A$. This suggests that experimental protocols may result in variability between cells in individual measurements of $S$ and $\gamma$. However, the *fitted* parameters $k$ and $d$ should not vary between cells. In particular, the *fitted* $d$ should correspond to the *real* $d$ which justifies using a measured PHD degradation rate as an initial estimate during optimization. Furthermore, we may be able to distinguish the effects of transient transfection separately from the effects due to the imaging protocol. Specifically, the imaging protocol should result in a correlation between $S^F$ and $\gamma^F$, such that the ratio $S^F/\gamma^F$ does not vary between cells, whereas transient transfection will lead to cell-to-cell variability in $S^R$ and consequent $S^F$ independently of $\gamma$.

**Acknowledgements**

JL was a recipient of a University of Liverpool studentship. VS was a recipient of a BBSRC David Phillips fellowship (BB/C520471/1). JB was a recipient of a BBSRC DTG studentship. AH has been funded by the Neuroblastoma Society and by the Alder Hey Oncology Fund ref 8098.

**Figures and Tables**

**Fig 1. Response of selected cells to transition from normoxic to hypoxic environment at $t = 0$.** Experimental HIF-1α data (red line); HIF-1α model output (blue line) and PHD model output (green line). Parameters were optimized to best fit the data as discussed in the text. In this fitting exercise the cells selected displayed bell-shaped transient dynamics.

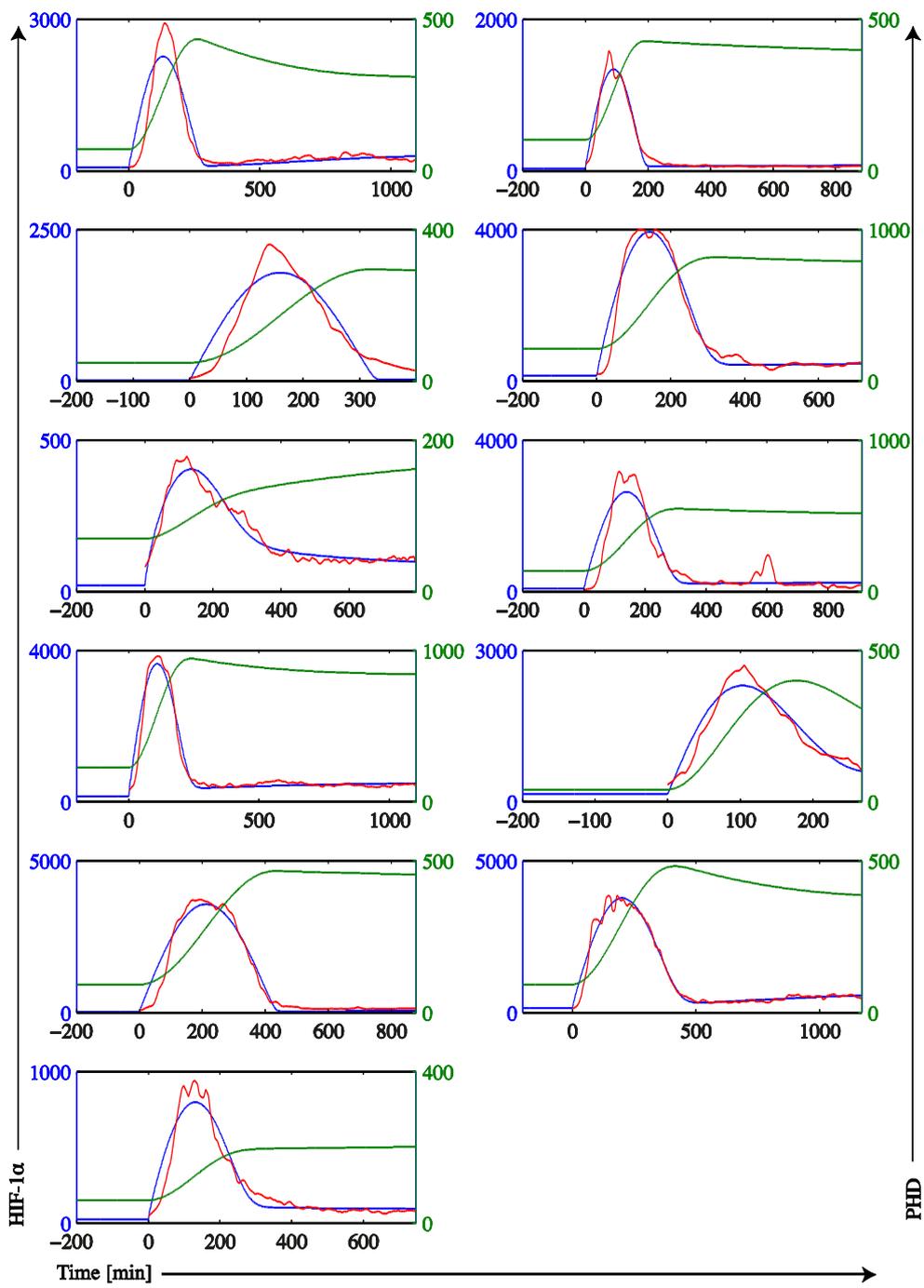

**Fig 2. Estimates for model parameters from free parameter optimization of bell-shaped dynamic data.** For each cell (x-coordinate 1-11), 50 initial sample parameter sets were chosen from which the search algorithm proceeded (see text for details). Final optimal parameter estimates for $S$, $\gamma$, $k$ and $d$ corresponding to best fits (minimum $\chi$ value) of the experimental data are indicated by circles on the figure. Final parameter estimates for searches which converged to a $\chi$ within 1 % of the minimum $\chi$ value are plotted as points on the figure. Horizontal red lines give median values ($k = 8.99 \times 10^{-4}$ min$^{-1}$; $d = 4.71 \times 10^{-4}$ min$^{-1}$) for optimal parameters.

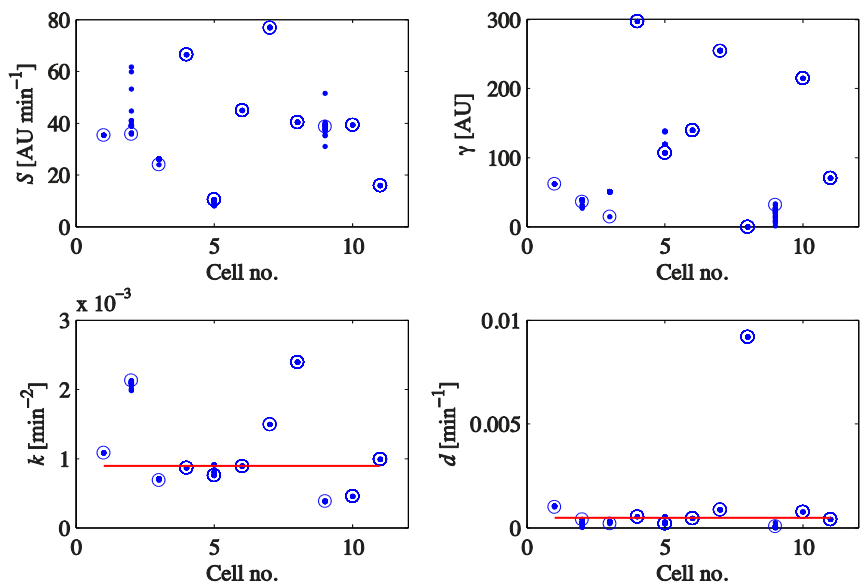

**Fig 3. Estimates for model parameters from constrained parameter optimization of all dynamic data.** Parameters were optimized for the full dataset of 39 cells using the constrained search algorithm to minimize the $\chi$ value. For each cell, final parameter estimates of searches that converged to a $\chi$ value within 1 % of the minimum $\chi$ were plotted, so long as the minimum $\chi$ value also corresponded to a good fit according to the envelope criteria. Different colors refer to different cell data as follows: *Experiment 1* (green), *Experiment 2* bell-shaped dynamics (blue) and *Experiment 2* non-bell-shaped dynamics (cyan).

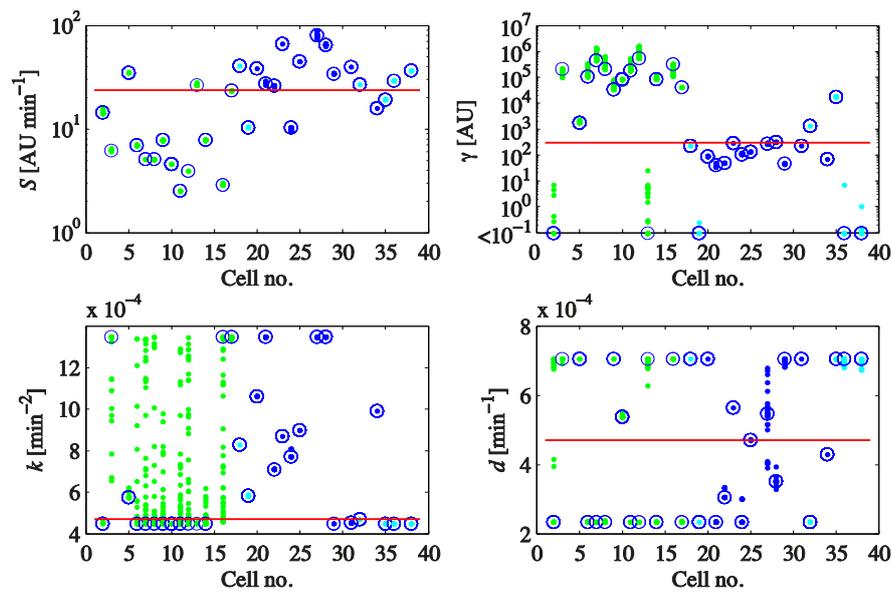

**Fig 4. Numerical simulations of 4-component model.** (A) In-silico hypoxic pulsing experiment. (B) Experimental results of hypoxic pulsing experiment.

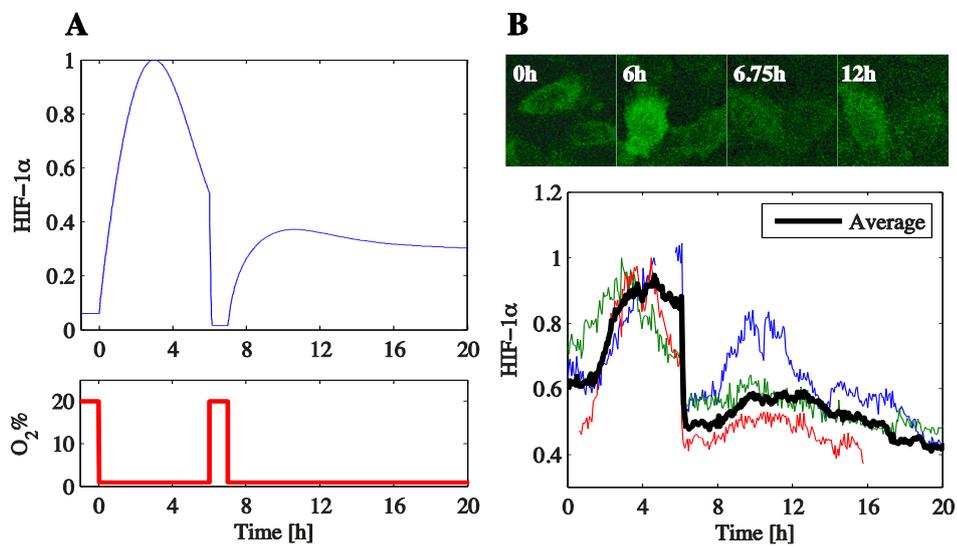

**Fig 5. Fitting oxygen data to reaction-diffusion model.** Experimental observations of oxygen life-time data within tumorsphere at 8 % external $O_2$ (A) and 3 % external $O_2$ (B). Scale bar is 100 μm. Measured internal steady-state oxygen concentration at line 1 (C) and line 2 (D) at 8 % external $O_2$, (blue) and 3 % external $O_2$ (red), distance non-dimensionalised on cell width at given line. Model solution with best-fit parameters for uptake and diffusion at 8 % external $O_2$ (green) and 3 % external $O_2$ (cyan).

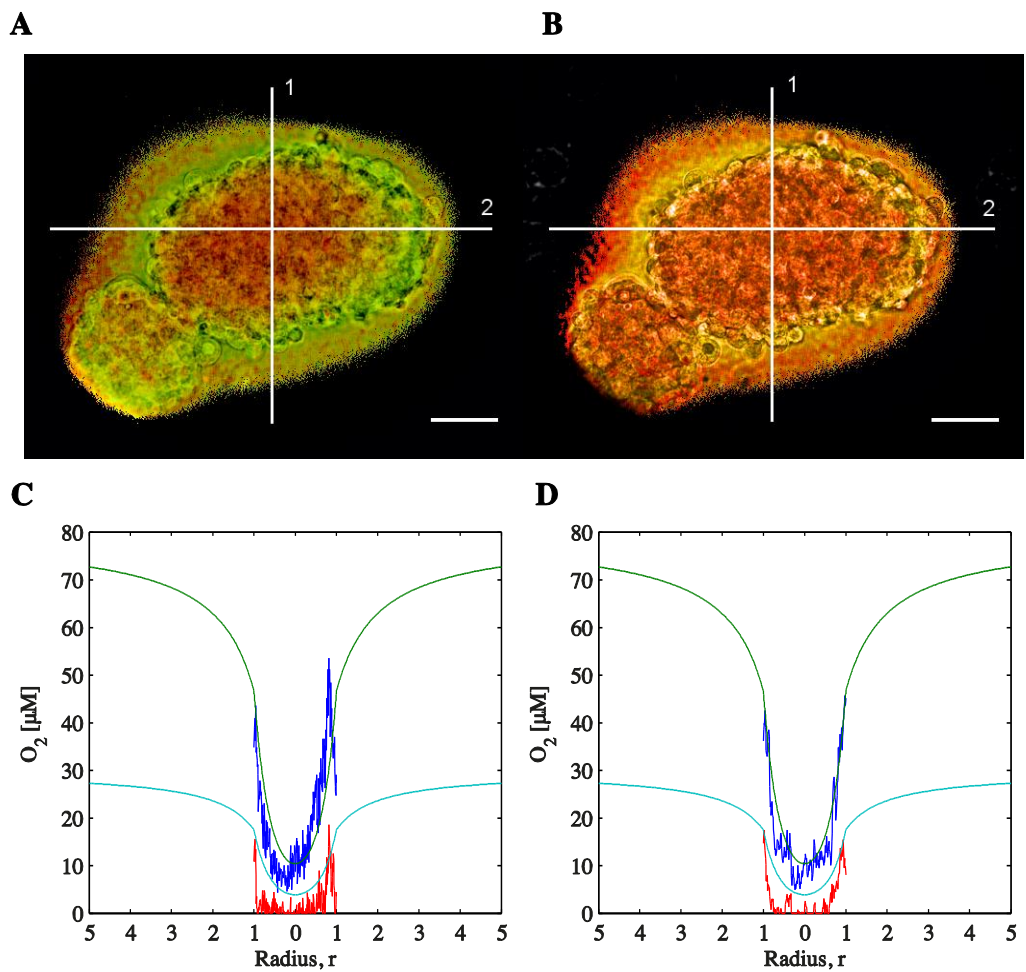

**Fig 6. Predictions of HIF-1α model coupled to spatial model of oxygen dynamics in sphere.** The upper cartoon depicts the coupled model. (A) Green/yellow sphere represent normoxic conditions, red/orange sphere represents hypoxic conditions, and cyan arrow indicates transition. (B) Sketch indicating oxygen temporal dynamics at an example spatial location, and (C) resultant HIF-1α dynamics. (D) Solution of oxygen reaction-diffusion equation within the sphere and resultant HIF-1α dynamics for when at $t = 0$ the external oxygen concentration is switched from 8 % to 3 %. The resultant internal oxygen (blue) and HIF-1α (black) dynamics are shown at indicated later times. Equilibrium solutions for the internal oxygen concentration in normoxia (green) and hypoxia (red) are also shown.

A

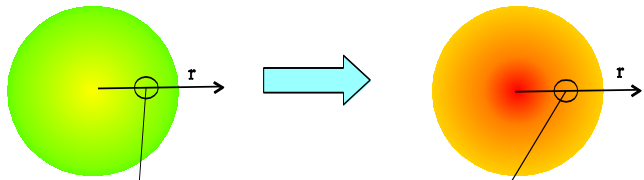

B

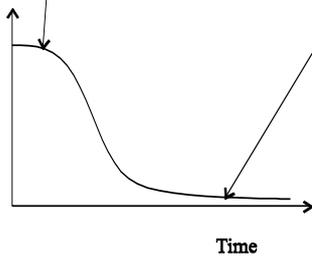

Oxygen

C

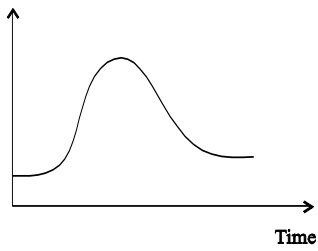

HIF-1α

D

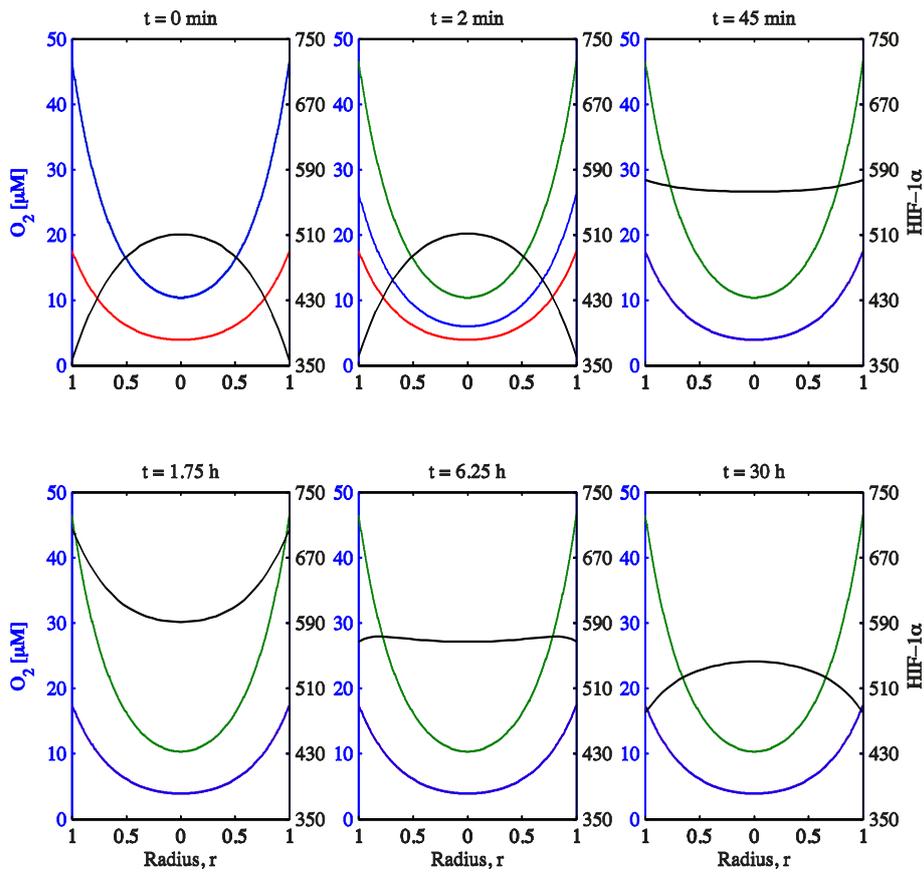

**Fig 7. Predictions of transcriptional targets in sphere.** Plots for HIF-1α, PHD2, PHD3 and internal oxygen for an external switch in oxygen from 20 % to 1 % at time $t = 0$ taken to occur at the boundary of a sphere 200 µm in radius. Within each plot, each curve represents the dynamics at a particular spatial position over time. The color-map ranges from blue to red indicating the radial distance from 0 µm (core) to 200 µm (boundary).

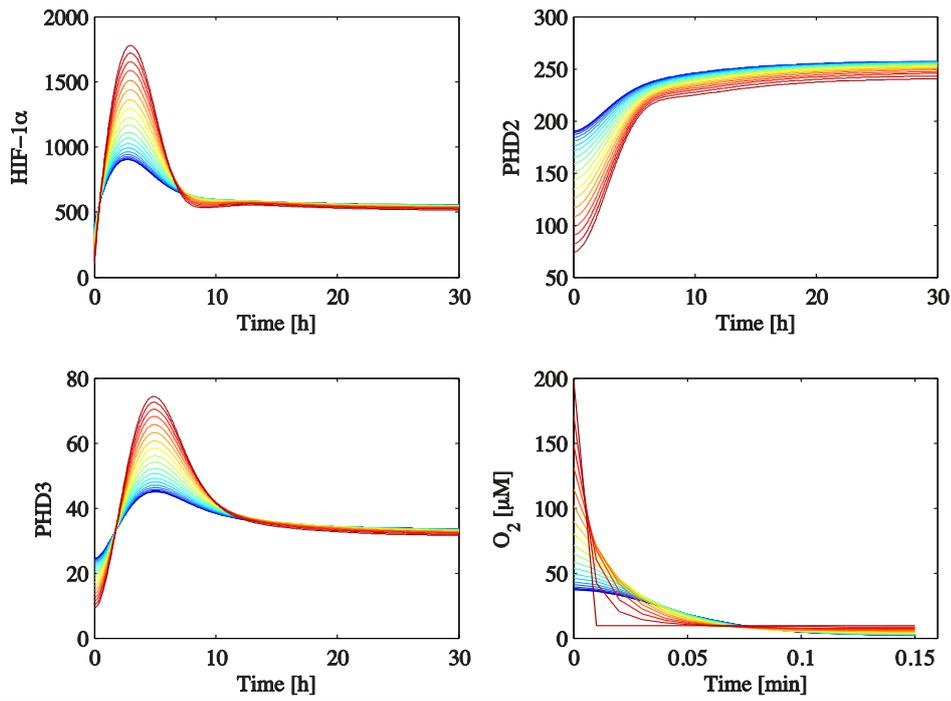

**Supplementary Figures**

**Fig S1 Constrained fit of bell data showing error envelope.**

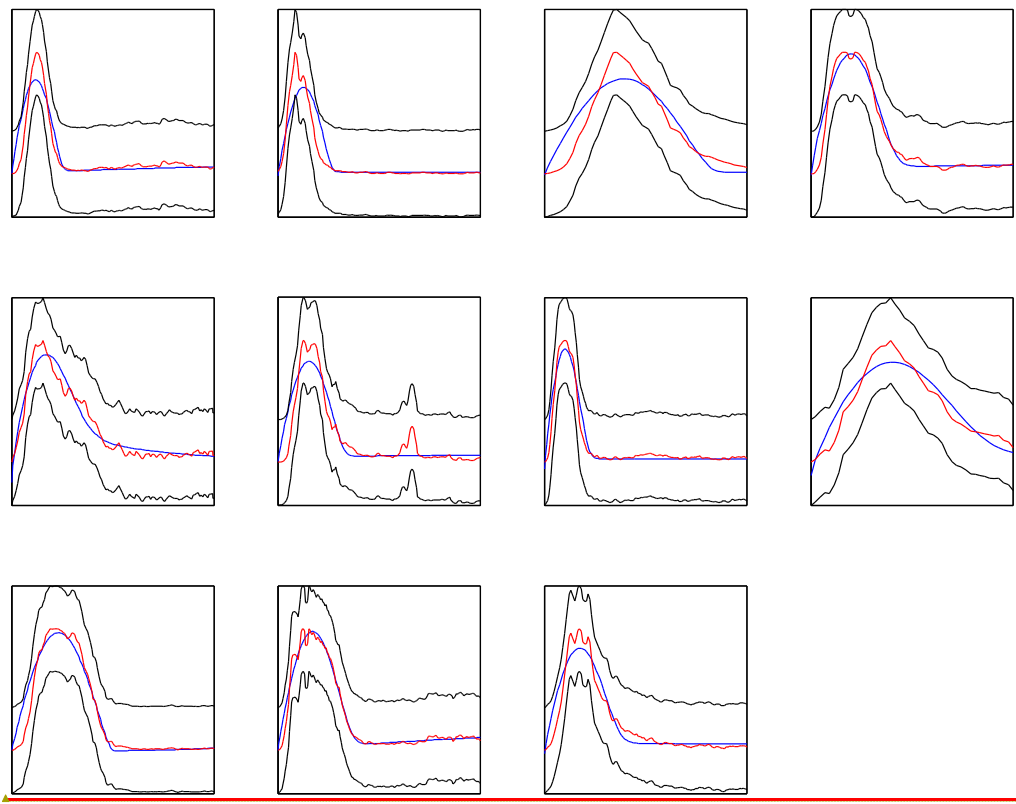

**Fig S2 Constrained fit of *Experiment 1* data showing error envelope.** Stars indicate bad fits according to the error envelope (see text for details).

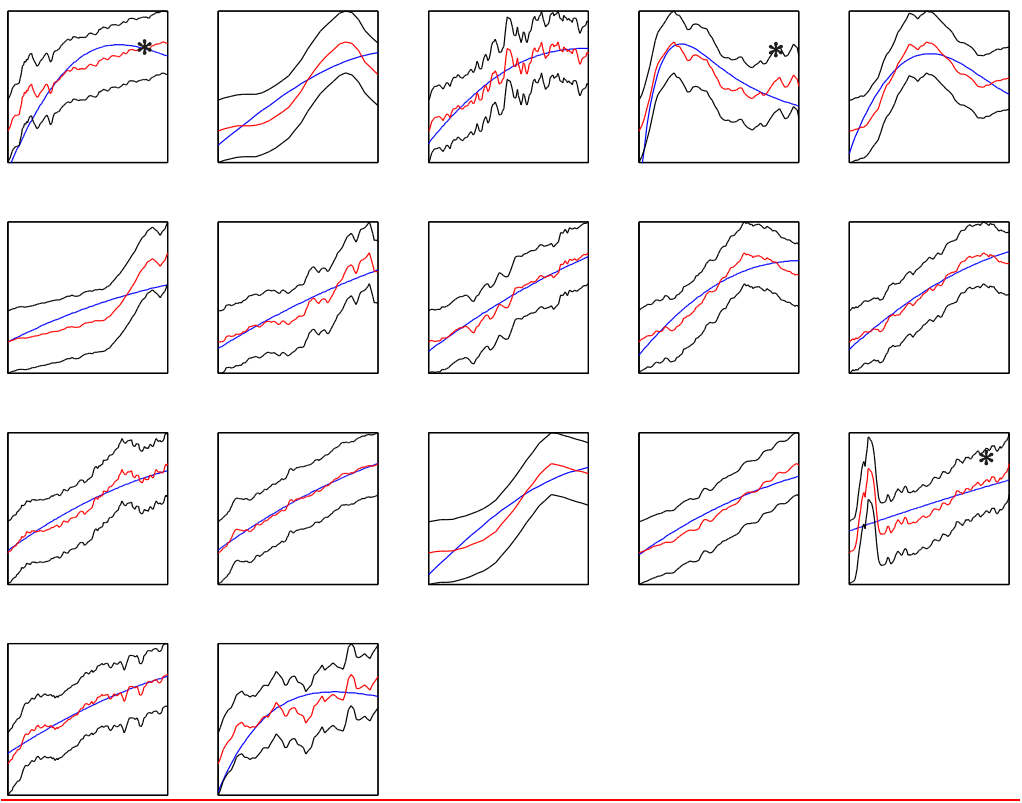

**Fig S3 Constrained fit of *Experiment 2* non-bell data showing error envelope.** Stars as Fig S2.

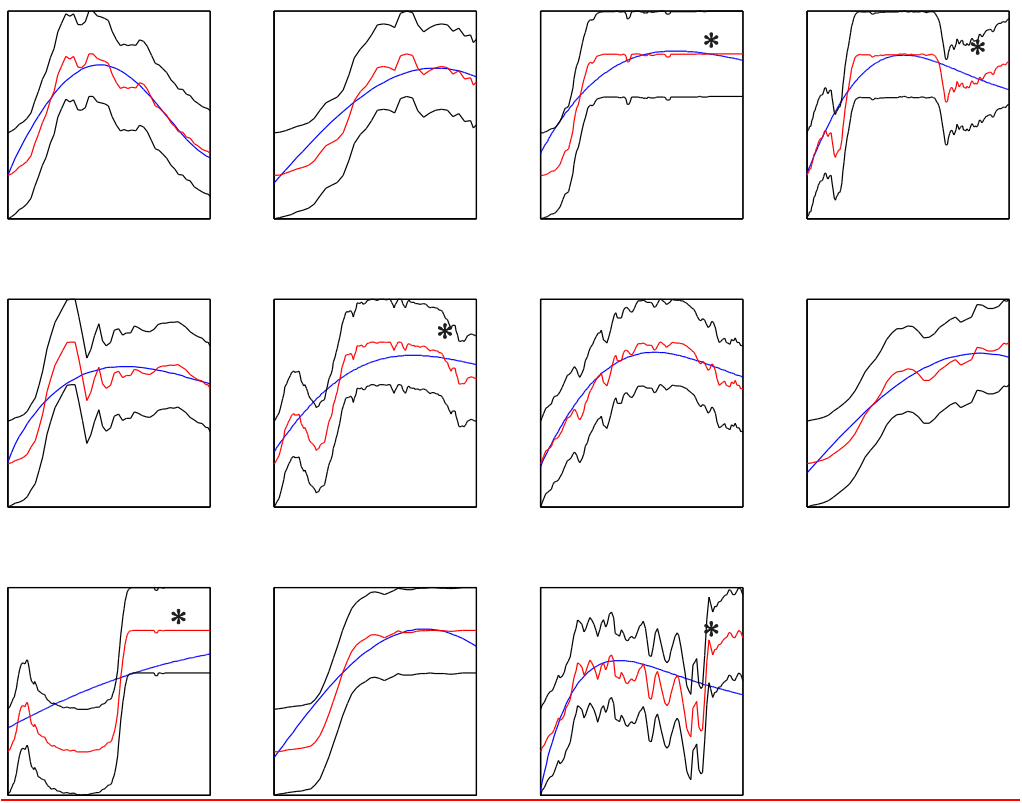